\documentclass[aps,pra,twocolumn,superscriptaddress,floats,epsfig,showpacs]{revtex4}

\usepackage{pstricks,pst-grad,color,graphics}
\usepackage{latexsym}
\usepackage{amsmath}
\usepackage{amssymb}
\usepackage{enumerate}
\usepackage{bbm}
\usepackage{psfrag}
\usepackage{epsfig}

\newcommand{\be}{\begin{equation}}
\newcommand{\ee}{\end{equation}}
\newcommand{\BE}{\begin{equation}}
\newcommand{\EE}{\end{equation}}
\newcommand{\eea}{\end{eqnarray}}
\newcommand{\bea}{\begin{eqnarray}}

\newcommand{\mean}[1]{\ensuremath{\langle{#1}\rangle}}

\newcommand{\eins}{\ensuremath{\mathbbm 1}}

\newcommand{\WW}{\ensuremath{\mathcal{W}}}

\newcommand{\NN}{\ensuremath{\mathcal{N}}}

\newcommand{\ketbra}[1]{\ensuremath{| #1 \rangle \langle #1 |}}

\newcommand{\ket}[1]{\ensuremath{|#1\rangle}}

\newcommand{\bra}[1]{\ensuremath{\langle#1|}}

\newcommand{\braket}[2]{\ensuremath{\langle #1|#2\rangle}}

\renewcommand{\vr}{\ensuremath{\varrho}}

\newcommand{\kommentar}[1]{}

\begin{document}
\title{Lower bounds on entanglement measures from incomplete information}

\author{O. G\"uhne}
\affiliation{Institut f\"ur Quantenoptik und Quanteninformation,
\"Osterreichische Akademie der Wissenschaften,
A-6020 Innsbruck, Austria}
\affiliation{Institut f\"ur theoretische Physik,
Universit\"at Innsbruck, 
Technikerstra{\ss}e 25, 
A-6020 Innsbruck, Austria}
\author{M. Reimpell}
\affiliation{Institut f\"ur Mathematische Physik, 
Technische Universit\"at Braunschweig, 
Mendelssohnstra{\ss}e 3, D-38106 Braunschweig, Germany}
\author{R.F. Werner}
\affiliation{Institut f\"ur Mathematische Physik, 
Technische Universit\"at Braunschweig, 
Mendelssohnstra{\ss}e 3, D-38106 Braunschweig, Germany}
\date{\today}

\begin{abstract}
How can we quantify the entanglement in a quantum state, 
if only the expectation value of a single observable is 
given?
This question is of great interest for the analysis of 
entanglement in experiments, since in many multiparticle 
experiments the state is not completely known. 
We present several results concerning this problem by considering 
the estimation of entanglement measures via Legendre transforms. 
First, we present a simple algorithm for the estimation of the 
concurrence and extensions thereof. Second, we derive an 
analytical approach to estimate the geometric measure of 
entanglement, if the diagonal elements of the quantum state
in a certain basis are known. Finally, we compare our bounds
with exact values and other estimation methods for entanglement 
measures.

\end{abstract}
\maketitle

\section{Introduction}
Entanglement is a key phenomenon in quantum information 
science and the quantification of entanglement is one of 
the major problems in the field. For this quantification 
many entanglement measures have been proposed \cite{plenio, 
eof, wei, hororeview, meyerwallach, vidalwerner, concurrence, 
osterloh1, mintertreview}. However, a central problem in 
most of these proposals is the actual calculation of a given 
measure: entanglement measures are typically defined via 
optimization procedures, which may consist of a maximization 
over certain protocols or the minimization over all decompositions 
of a state into pure states.
For some remarkable cases it happens that such minimizations can 
be performed analytically \cite{wootters1, terhalvollbrecht, 
rungta, osterloh2}, however, in the general case these problems 
are not solved. Therefore, to take a realistic point of view, 
one can try to estimate entanglement measures, and many 
proposals for the estimation of entanglement measures have been 
presented \cite{cab1, cab2, mintert1, vicente1, datta}.

In an experimental setting, the situation becomes even more 
complicated: since quantum state tomography for multiparticle
systems requires an exponentially increasing effort, the state
is often not completely known. Typically, one measures so-called
entanglement witnesses, special observables, for which a negative
expectation value signals the presence of entanglement \cite{horo, 
terhal, mohamed, kiesel, haeffner, lu, tothguehne}. In this sense, 
entanglement witnesses allow to detect entanglement, but the question 
arises, whether they also allow to quantify entanglement.
This question has been addressed from several perspectives \cite{horoold,wolf, 
brandao, plenio2, mintert2} and in Refs.~\cite{wir, eisert} a general 
recipe for this problem was found. There it has been shown how one 
can derive the optimal lower bound on a generic entanglement 
measure from the expectation value of a witness or another 
observable. The estimate uses Legendre transforms to give 
lower bounds on the convex entanglement measure, and the main
task in this scheme is to compute the Legendre transform
of a given entanglement measure.

In this paper, we extend this method into several directions. 
First, we derive a simple algorithm for the calculation of the 
Legendre transform for the concurrence \cite{concurrence} and 
extensions thereof. Then, we present analytical results for the
Legendre transform for certain witnesses for the geometric measure
of entanglement \cite{wei}. Finally, we discuss examples and compare 
our results to other methods for entanglement estimation. But before
presenting the new results, let us shortly review the method presented 
in Refs.~\cite{wir,eisert}.

\section{The method}
Let us consider the following situation: in an experiment, 
an  entanglement witness $\WW$ has been measured and the 
mean value $\mean{\WW}=Tr(\vr \WW)=w$ has been found. 
The task is now 
to derive from this single expectation value a quantitative
statement about the entanglement present in the quantum 
state. In our case, we aim at  providing a lower bound on 
the entanglement inherent in the  state $\vr$. That is, 
we are looking for statements like
\be
Tr(\vr \WW)=w \;\; \Rightarrow \;\; E(\vr) \geq f(w),
\label{task}
\ee
where $E(\vr)$ denotes an arbitrary convex and continuous 
entanglement measure. We do not specify it at this point 
further. Naturally, we aim to derive an optimal bound $f(w)$ 
and an estimate is optimal, if there is a state $\vr_0$ with 
$Tr(\vr_0 \WW)=w$ and $E(\vr_0) = f(w).$

In order to derive such lower bounds, let us consider 
the so-called Legendre transform of $E$ for the witness 
$\WW$, defined via the maximization
\be
\hat{E}(\WW) = \sup_\vr \{Tr(\WW \vr) - E(\vr)\}.
\label{legtrafo}
\ee
As this is defined as the maximum over all $\vr,$ we have 
for any fixed $\vr$ that $\hat{E}(\WW) \geq  Tr(\WW \vr) - E(\vr),$ 
hence
\be
E(\vr) \geq Tr(\WW \vr) - \hat{E}(\WW),
\label{firstbound}
\ee
which is known as Fenchel's inequality or Young's inequality.
The point is that the first term on the right hand side are 
the given measurement data, while the second term can be computed. 
Therefore, a measurable bound on $E(\vr)$ has been obtained.

In order to improve this bound, note that knowing the data
$Tr(\vr \WW)=w$ is, of course, equivalent to knowing 
$Tr(\vr \lambda \WW)= \lambda w$ for any $\lambda.$ Therefore, 
we can optimize over all $\lambda$ and obtain
\be
E(\vr) \geq 
\sup_\lambda \{ \lambda Tr(\WW \vr) - \hat{E}(\lambda\WW)\}.
\label{secondbound}
\ee
This is a better bound than Eq.~(\ref{firstbound}) and, as 
we will see, already the optimal bound in Eq.~(\ref{task}).

\begin{figure}[t]
\includegraphics[width=0.9\columnwidth]{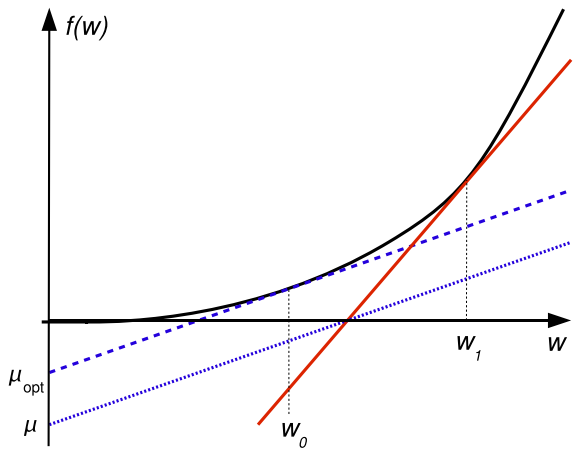}
\caption{Geometrical interpretation of the Legendre transform. 
The dotted line is a general affine lower bound, and the dashed
line is an optimized bound with the minimal $\mu.$ By varying
the slope, one finds an affine lower bound, which is tight
for a given $w_1$ (solid line).
See text for details. 
}
\end{figure}

For our later discussion, it is important to note that this
estimation has a clear geometrical meaning [see Fig.~(1)].
As $E(\vr)$ is convex, the minimal $E(\vr)$ compatible with 
$Tr(\vr \WW)=w,$ denoted by $f(w)$ is convex, too \cite{remark1}. 
Let us consider a generic affine lower bound $g(w)= \lambda w - \mu$ 
on $f(w),$ i.e., $f(w) \geq \lambda w - \mu.$ We have 
$\mu \geq \lambda w -f(w),$ and in order to make the bound 
for a fixed $\lambda$ as good as possible, we have to choose 
$\mu$ as small as possible. This leads to 
$\mu_{\rm opt}=\sup_w \{\lambda w - f(w)\}$ which is exactly the 
optimization in Eq.~(\ref{legtrafo}).

For a fixed slope $\lambda$ we obtain by this method an affine 
bound (characterized by $\mu_{\rm opt}$) which is already optimal
for a certain value $w_0$, as it touches $f(w)$ in this point.
For any other mean value $w$ it delivers a valid, but not 
necessarily optimal bound. To obtain the optimal bound for 
any given $w_1$ we have to vary the slope $\lambda.$ This corresponds 
to the optimization  in Eq.~(\ref{secondbound}). Since $f$ is convex, 
we obtain for each $w$ the tight linear bound, showing that this
optimization  procedure gives indeed the best possible bounds in 
Eq.~(\ref{task}).
 
Three remarks are in order at this point. First, it was not needed
that $\WW$ is an entanglement witness, we may consider an arbitrary
observable instead. Second, we can also consider a set of observables
$\vec{\WW} = \{\WW_1, ..., \WW_n\}$ at the same time. We just have to 
introduce a vector $\vec{\lambda} =\{\lambda_1,...,\lambda_n\}$
and replace $\lambda \WW$ by $\sum_k \lambda_k \WW_k,$ then all
formulas remain valid. Finally, also for a non-convex
$E(\rho)$ the method delivers valid bounds, however, then it is not 
guaranteed that the bounds are the optimal ones.

In any case, the method relies vitally on the ability to compute the 
Legendre transform in Eq.~(\ref{legtrafo}). The difficulty of 
this task clearly depends on the witness $\WW$ and on the measure 
$E(\vr)$ chosen. At first sight, the task may seem hopeless, 
as the calculation of $E(\vr)$ for mixed states is for many measures
already impossible. 

For instance, a large class of entanglement measures is defined 
via the convex roof construction. For that, one first defines the 
measure  $E(\ket{\psi})$ for pure states, and then defines for 
mixed states
\be
E(\vr) = \inf_{p_k, \ket{\phi_k}} \sum_k p_k E(\ket{\phi_k}),
\label{crc}
\ee
where the infimum is taken over all possible decompositions of $\vr,$
i.e., over all $p_k$ and $\ket{\phi_k}$ with 
$\vr=\sum_k p_k \ketbra{\phi_k}.$ Clearly, this infimum is very 
difficult to compute.

However, for the computation of the Legendre transform, this is not 
relevant: as one can easily prove (see Ref.~\cite{wir} for details), 
for convex roof measures the maximization has to run only over pure 
states
\be
\hat{E}(\WW) = \sup_{\ket{\psi}} 
\{ 
\bra{\psi}\WW\ket{\psi} - E(\ket{\psi})
\},
\ee
which simplifies the calculation significantly. In fact, this shows
that for convex roof measures, which are by construction rather 
difficult to compute, the Legendre transform is rather simple to 
compute.

In Ref.~\cite{wir} we have considered the entanglement of formation
and the geometric measure of entanglement as entanglement measures,
which are both convex roof measures. We have provided simple algorithms 
for the calculation of the Legendre transform, and for special witnesses,
we calculated the Legendre transform of the geometric measure also 
analytically.
In this paper, we will consider the concurrence and extensions thereof, and we 
will also present new analytical results for the geometric measure.

Finally, it should be noted that the optimization in Eq.~(\ref{secondbound})
can be completely skipped. Any $\lambda$ delivers already a valid bound. 
However, the optimal $\lambda$ can easily be found numerically.

\section{The concurrence}

As a first entanglement measure, let us discuss the 
concurrence. This quantity is defined for pure states
as \cite{concurrence}
\be
E_C(\ket{\psi}) = \sqrt{2[1-Tr(\vr_A^2)]},
\label{concdef}
\ee
where $\vr_A$ the reduced state of $\ket{\psi}$ 
for Alice. For mixed states this definition is extended
via the convex roof construction in Eq.~(\ref{crc}). 
For the special case of two qubits, the concurrence is a 
monotonous function of the entanglement of formation, 
moreover, the minimization in the convex roof construction 
can be explicitly performed \cite{wootters1}. In the general 
case, it is not so directly connected to the entanglement 
of formation. The concurrence is an entanglement monotone 
\cite{vidal}, but it does not fulfill all desired axioms 
for an entanglement measure, e.g.~it is not additive. 

Here, we want to calculate the Legendre transform of $E_C$ 
for a generic witness $\WW.$ Similar as it has been done in  
Ref.~\cite{wir} for the entanglement of formation, we will 
construct an iterative algorithm for the optimization, 
which converges to the maximum as a  fixed point. 

First, as the concurrence is defined via the convex roof, 
it suffices in Eq.~(\ref{legtrafo}) to optimize over pure 
states only. Then, using the fact that for $x \in [0;1]$
\be
\sqrt{x} = 
\inf_{\alpha \in [0;1]} \{\frac{x}{2{\alpha}} + \frac{{\alpha}}{2}\},
\ee
we can rewrite the Legendre transform as
\be 
\hat{E}_C(\WW) = 
\sup_{\ket{\psi}}\;\; \sup_{\alpha}\{ \bra{\psi}\WW\ket{\psi} 
-  \frac{1-Tr(\vr_A^2)}{\sqrt{2}{\alpha}} - \frac{{\alpha}}{\sqrt{2}}\}.
\ee
The idea is to write this maximization as an iteration that 
optimizes $\alpha$ and $\ket{\psi}$ in turn. For a fixed 
$\ket{\psi}$ we perform the maximization over $\alpha$ analytically 
and similarly we can find the optimal $\ket{\psi}$ for a fixed 
$\alpha.$ Concerning the first step, note that for 
a fixed $\ket{\psi}$ the optimal $\alpha$ is simply given by
\be
\alpha= \sqrt{1-Tr(\vr_A^2)}.
\ee
Concerning the second step, if $\alpha$ is fixed, 
we have essentially to solve an optimization 
problem like
\be
\sup_{\ket{\psi}} \{ \bra{\psi} \tilde{\WW} \ket{\psi} - [1-Tr(\vr_A^2)]\},
\ee
where $\tilde{\WW}$ is proportional to the original witness $\WW.$ 

Here, the second term is nothing but the q-entropy, 
investigated in detail by Havrda, Charvat, Dar\'oczy 
and Tsallis \cite{havrda, daroczy, wehrl, tsallis},
\be
S_q (\vr) = \frac{1-Tr(\vr^q)}{q-1}
\ee
for the case $q=2.$ Therefore, we can try to write $S_q (\vr_A)$ 
as an infimum via the Gibbs principle, similarly as it has been 
done for the von Neumann entropy in Ref.~\cite{wir}.

So we make the ansatz
\bea
S_2(\vr)&=&\inf_H \{ Tr(\vr H) - F_2(H)\}
\label{opti1}
\\
F_2(H)&=& \inf_\vr \{Tr(\vr H) - S_2(\vr)\}
\label{opti2}
\eea
For the case of the von Neumann entropy these formulas
just express the Gibbs variational principle, where $F$
is the free energy, and the inverse temperature was set 
to $\beta=1.$

We are mainly interested in the first minimization and 
have to compute $F_2$ and the $H,$ where the minimum 
is attained. The point is that the second minimization 
has been solved already by Guerberoff and Raggio \cite{guerberoff} 
and the unique thermal state minimizing Eq.~(\ref{opti2})
for an arbitrary Hamiltonian $H$ and consequently $F_2$ is 
known. For the first minimization, it remains to find the 
Hamiltonian for which the given state is the thermal state.

In order to do this in practice, let us first recall the results
of Ref.~\cite{guerberoff}. For the case $q=2$ and $\beta>0$ the 
results of this reference state the following:

Let $H$ be a Hamiltonian with ground state energy 
$\varepsilon_-$, ground state degeneracy $g_-$ and 
let the energy of the first excited state be 
$\varepsilon_-^*.$ Define
\be
\beta^+=\frac{2}{g_-(\varepsilon_-^*-\varepsilon_-)} > 0  
\;\;\mbox{ and }\;\;
t^+=\frac{1}{(\varepsilon_-^*-\varepsilon_-)} > 0,
\ee
and the monotonically increasing function 
$\beta(t):[0,t^+]\rightarrow [0,\beta^+]$
as
\be
\beta(t)=\frac{2t}{Tr([\eins-t(H-\varepsilon_-\eins)]_\oplus)},
\label{betadef}
\ee
where $[X]_\oplus$ denotes the positive part of the operator
$X.$ Let $\tau$ be the inverse function to $\beta(t).$

Then, for $\beta < \beta^+$ the unique thermal state is 
given by
\be
\vr = \NN ([\eins - \tau(\beta)(H-\varepsilon_-\eins)]_\oplus)
\label{thermal}
\ee
where $\NN$ denotes the normalization, and for $\beta \geq \beta^+$
the thermal state is just the normalized projector onto the eigenspace
corresponding to the lowest energy $\varepsilon_-.$ From this, it is 
clear that $\vr$ and $H$ are diagonal in the same basis. 

Coming back to the minimizations in Eqs.~(\ref{opti1}, \ref{opti2})
let us assume that a density matrix $\vr$ with decreasing eigenvalues 
$\lambda_i$ ($i=1,...,N$) is given, and the task is to compute the 
corresponding Hamiltonian $H$ with increasing eigenvalues $E_i.$ 
Without loosing generality, we can choose $E_1=0.$ 

Let us first consider the case that $\vr$ is non degenerate and 
has full rank. Then, Eq.~(\ref{thermal}) implies that the 
eigenvalues $E_i$ of $H$ have to fulfill $1-\tau(\beta)E_i > 0$ 
for all $i.$ Since we are considering the case $\beta=1,$
we have for $\tau_0=\tau(\beta=1)$ due to Eq.~(\ref{betadef})
that
$
\tau_0 = [Tr(\eins - \tau_0 H)]/2.
$
{From} Eq.~(\ref{thermal}) is follows first that 
$\lambda_1=\NN=1/Tr(\eins-\tau_0 H)=1/(2\tau_0)$, 
and then
$
\lambda_i = \lambda_1 (1-\tau_0 E_i)
=
\lambda_1[1-E_i/(2 \lambda_1)],
$
hence 
\be
E_i = 2 (\lambda_1-\lambda_i).
\ee
For this Hamiltonian, we have also
\be
F_2(H)= 2 \lambda_1 - \sum_i \lambda_i^2 - 1.
\ee
As a remark, first note that this solution fulfills
$1-\tau(\beta)E_i=1-(\lambda_1-\lambda_i)/\lambda_1 > 0,$ 
as requested at the beginning. Second, it delivers 
$\beta^+=1/(\lambda_1-\lambda_2) > \beta = 1,$ justifying 
the ansatz in Eq.~(\ref{thermal}). Finally, it is easy to 
see that if $\vr$ is not of full rank or degenerated, 
the recipe for above works also and delivers a correct 
solution.

In summary, we can write the problem of the computation 
of the Legendre transform for the concurrence as:
\begin{widetext}
\be
\hat{E}_C(\WW) = 
\sup_{\ket{\psi}}
\;\; 
\sup_{H}
\;\; 
\sup_{\alpha}
\Big\{ 
\bra{\psi}\WW\ket{\psi} 
-\frac{1}{\sqrt{2}{\alpha}} 
\big[\bra{\psi}(H \otimes \eins)\ket{\psi} - F_2(H)\big]
- \frac{{\alpha}}{\sqrt{2}}
\Big\}.
\ee
\end{widetext}
If $\ket{\psi}$ and $H$ are fixed, we can compute the optimal 
$\alpha$ as shown in the beginning. If $\ket{\psi}$ and $\alpha$
are fixed, we can determine the optimal $H$ as above. Finally, 
if $H$ and $\alpha$ are fixed we choose $\ket{\psi}$ as the eigenvector
corresponding to the maximal eigenvalue of 
$\WW-(H \otimes \eins)/(\sqrt{2}{\alpha}).$ 
Therefore, we have an iterative optimization, which delivers an
monotonously increasing sequence of values for the Legendre 
transform, with the actual value as a fixed point. 

This algorithm can be implemented with few lines of code, 
examples will be discussed in Section V. It should be noted 
that in principle we cannot prove that the algorithm converges 
always to the global optimum, as the final fixed point may depend
on the starting values of $\ket{\psi}$, $H$ and $\alpha,$ leading 
to an overestimation of the entanglement. In practice, however, 
the convergence behavior is quite good natured and the algorithm 
delivers a well suited tool for obtaining sharp bounds on the 
concurrence.

Finally, let us add that with the same method also extensions of the 
concurrence may be treated. For instance, one may consider quantities
for pure states as $E_q(\ket{\psi}) = S_q(\vr_A).$ With the convex 
roof extension, these are also entanglement monotones \cite{vidal},
and using the results of Ref.~\cite{guerberoff}, the Legendre 
transform can be calculated in a similar manner as above.

In addition, there are several multipartite entanglement 
monotones, which can be seen as multipartite extensions 
of the concurrence \cite{meyerwallach, brennen, mintertreview}.
For instance, for $N$-qubit states one can define the Meyer 
Wallach measure
\be
E_{MW}(\ket{\psi})= 
2 \big[1-\frac{1}{N}\sum_{k=1}^{N}Tr(\vr_k^2)\big],
\ee
where the $\vr_k$ are all reduced one-qubit states. Clearly,
due to the similar structure as in Eq.~(\ref{concdef}) the 
methods from above can also be used to compute the Legendre 
transformation for these measures.

\section{The geometric measure}

As the second entanglement measure, let us consider the 
geometric measure of entanglement $E_G$ \cite{wei}. This 
is an entanglement monotone multipartite systems, quantifying 
the distance to the separable states. The geometric measure 
is defined for pure  states as 
\be
E_G(\ket{\psi}) = 1- \sup_{\ket{\phi}=\ket{a}\ket{b}\ket{c}...}
|\braket{\phi}{\psi}|^2,
\ee
i.e., as one minus the maximal overlap with pure fully separable 
states, and for mixed states via the convex roof construction.

The geometric measure is a {\it multipartite} entanglement 
measure, as it is not only a summation over bipartite entanglement 
properties. Despite of its abstract definition, it has turned
our that $E_G$ can be used to quantify the distinguishability
of multipartite states by local means \cite{hayashi}. 
As the geometric measure is one of the few measures for 
multipartite systems, which have a reasonable operational 
meaning and are at the same time proved to fulfill all of 
the conditions for entanglement monotones, it has been 
investigated from several perspectives, for 
instance it has been used to study multipartite entanglement 
in condensed matter systems \cite{geospin}.

In 
Ref.~\cite{wir} the problem of calculating the Legendre transform
for the geometric measure was already considered and the following 
results were obtained:

First, an iterative algorithm (in the same 
spirit as the algorithm for the concurrence in the previous section)
has been derived for calculating the Legendre transform of arbitrary 
witnesses. Second, for the important case that the witness is of 
the form $\WW = \alpha \eins - \ketbra{\chi}$ (or, equivalently 
$\WW = \ketbra{\chi}$) an analytic formula of the Legendre 
transform has been derived, reading, 
\begin{align}
&\hat{E}_G(r\WW) = 
\nonumber
\\
&
\left\{ 
\begin{array}{l}
r\alpha  \hfill \mbox{ for } r \geq 0,
\\
\big[\sqrt{(1-r)^2+4rE_G(\ket{\chi})} + 2\alpha r - r -1
\big]/2
\;\;\mbox{ for } r < 0.
\end{array}
\right.
\label{singlefid}
\end{align}
Here, we want to generalize this result by determining
analytically $\hat{E}_G(\WW)$ for the case that $\WW$ 
is diagonal in some special basis, e.g. the GHZ-state 
basis. We first consider the special case of the GHZ 
state basis, then the formula for the general case 
can directly be written down.

So let us assume that
\be
\WW = \sum_{i=1}^{2^N} \lambda_i \ketbra{GHZ_i}
\ee
is the witness for an $N$ qubit system, where 
$\ket{GHZ_i}=(\ket{x^{(1)}...x^{(N)}} \pm 
\ket{y^{(1)}...y^{(N)}})/\sqrt{2}$ with 
$x^{(i)}, y^{(i)} \in \{0,1\}$ and 
$x^{(i)}\neq y^{(i)}$ is the GHZ-state basis.
Without loosing generality we assume that the 
$\lambda_i$ are decreasingly ordered, i.e., $\lambda_1 
\geq \lambda_2 \geq ...\geq\lambda_{2^N},$ but 
not necessarily positive.

The Legendre transform is given by
\bea
\hat{E}_G(\WW) &=& 
\sup_{\ket{\psi}} 
\sup_{\ket{\phi}=\ket{a}\ket{b}\ket{c}...}
\bra{\psi} \big[ \WW + \ketbra{\phi}\big]\ket{\psi} - 1
\nonumber
\\
&=&
\sup_{\ket{\phi}=\ket{a}\ket{b}\ket{c}...} 
\big\Vert
\big[ \WW + \ketbra{\phi}\big]
\big\Vert
-1,
\eea
where $\Vert X \Vert$ denotes the maximal eigenvalue of the 
operator $X.$ In order to compute this maximal eigenvalue, we 
write the operator $[ \WW + \ketbra{\phi}]$ in the GHZ basis. 
$\WW$ is diagonal there and since the maximal overlap between 
the fully separable state $\ket{\phi}$ and any of the GHZ 
states is $1/2$ (i.e.,~the geometric measure for GHZ states is 
$1/2$ \cite{wei}), the matrix representation of $\ketbra{\phi}$ 
has matrix elements with absolute values not larger than $1/2.$

Our claim is now that the optimal choice of $\ket{\phi}$ is 
to take $\ketbra{\phi}$ as a $2\times 2$ matrix 
with all entries $1/2$ and acting on the two-dimensional space 
corresponding to the largest eigenvalues $\lambda_1$ and 
$\lambda_2.$ 
To prove that this choice is really optimal, 
we show that the above mentioned choice is also optimal 
if we consider the more general class of all $\ket{\phi},$ 
which have an overlap smaller or equal $1/2$ with all the GHZ 
states, but which are not necessarily product states. 

We prove it by contradiction. 
Let us assume a different optimal solution 
$\ket{\phi}$ and a corresponding eigenvector to 
the maximal eigenvalue $\ket{\psi}.$   
The vectors can be written as 
$\ket{\phi}=\sum_k \alpha_k \ket{GHZ_k}$
and $\ket{\psi}=\sum_k \beta_k \ket{GHZ_k}.$
We assume without loosing generality that 
$0 < |\alpha_1|^2 < 1/2$ and $ 0 < |\alpha_3|^2 < 1/2 .$ 
The function to maximize is given by
\be
\hat{E}_G(\WW) = 
\big| \sum_k \alpha_k^* \beta_k \big|^2 
+ \sum_k |\beta_k|^2 \lambda_k -1.
\ee
Since we are interested in the maximum, we can, without 
restricting generality assume that $\alpha_k$ and $\beta_k$ 
are real and positive. The interesting terms for the discussion 
are
\be
X= \alpha_1 \beta_1 + \alpha_3 \beta_3 ;  
\;\;\;
Y= \beta_1^2 \lambda_1 +  \beta_3^2 \lambda_3.
\ee
$X$ is a scalar product, which is maximal if the vectors 
$(\alpha_1, \alpha_3)$ and $(\beta_1, \beta_3)$ are parallel.
For given values of $(\beta_1, \beta_3)$ this may be 
prohibited by the constraint $\alpha_i^2 \leq 1/2.$ Then, 
however, it is clearly optimal to take an 
$(\alpha_1, \alpha_3)$ at the border of the domain, 
which has for one $\alpha_i=1/2,$ leading to a 
contradiction to the assumption on the form of $\ket{\phi}.$
Otherwise, we can choose the two vectors parallel, and 
$(\alpha_1, \alpha_3)$ is not at the border. Then, 
however, we can enlarge $Y$ by increasing $\beta_1$ in 
$(\beta_1, \beta_3)$ (and, simultaneously $\alpha_1$ in 
$(\alpha_1, \alpha_3)$ in order to keep the vectors parallel). 
This leads to another contradiction concerning the optimality
of $\ket{\psi}.$

Note that the class of the considered $\ket{\phi}$
is strictly larger than the class of product vectors, 
since not for any pair of GHZ states $\ket{GHZ_1}$ 
and $\ket{GHZ_2}$ we can find a product vector $\ket{\phi},$ 
such that $\ket{\phi}$ has an overlap of $1/2$ with both of 
the $\ket{GHZ_i}.$ However, we obtain an upper bound on the 
Legendre transform from this ansatz, which can be used for 
a valid lower bound on the entanglement measure. Further, 
one can check whether this bound is tight by direct 
inspection of  $\ket{GHZ_1}$ and $\ket{GHZ_2}$ afterwards.

Having shown that the simplest choice of $\ket{\phi}$ is 
optimal, the  calculation of the Legendre transform reduces 
to a calculation of eigenvalues of a $2\times 2$-matrix, and 
we have: 
\bea
\hat{E}_G(\WW) &\leq& 
\Big\Vert
\left[
\begin{array}{c c}
\lambda_1 + 1/2 & 1/2\\
1/2 & \lambda_2 + 1/2
\end{array}
\right]
\Big\Vert
-1
\nonumber
\\
&=& \frac{\lambda_1 + \lambda_2  - 1}{2} 
+ \frac{1}{2}\sqrt{(\lambda_1-\lambda_2)^2-1}.
\eea

The generalization to other states besides GHZ states 
is straightforward: if the overlap is bounded by 
some other number (e.g. 1/4 for four-qubit cluster 
states), we only have to calculate the eigenvalues 
of some larger matrix (e.g. a $4\times 4$-matrix 
for four-qubit cluster states), in order to derive 
an analytical upper bound on $\hat{E}_G(\WW).$ 
We can summarize: 

{\bf Observation.} Let 
$\WW = \sum_{i=1}^{2^N} \lambda_i \ketbra{\psi_i}$
be an operator, where for all eigenvectors 
$\ket{\psi_i}$ the overlap with fully separable states 
is bounded by $1/k, k\in \mathbbm{N}.$ 
Then the Legendre transform is bounded by 
\be
\hat{E}_G(\WW) \leq
\big\Vert
\left[
X
\right]
\big\Vert
-1
\label{anageo}
\ee
where $X$ is a $k\times k$-matrix with the entries 
$\lambda_i + 1/k$ on the diagonal, and offdiagonal 
entries $1/k.$ The question whether this bound is 
the exact value, can be decided by direct inspection 
of the $\ket{\psi_i}.$

This Observation allows for a simple calculation of a 
lower bound on $E_G$ if the fidelities of the basis
states $\ket{\psi_i}$ are known. 
First, the estimation is much simpler and 
faster compared with the iteration algorithm for arbitrary 
witnesses, since the optimization runs only over the $k$ 
largest eigenvalues $\lambda_i$ of the possible witnesses 
$\WW = \sum_i \lambda_i \ketbra{\psi_i}$.
For the iteration algorithm, it would be necessary to consider
all witnesses $\WW = \sum_i \lambda_i \ketbra{\psi_i}$, which
amounts to a variation over $2^N$ parameters $\lambda_i$. 
Second, the bounds on $E_G$ may become significantly 
better, compared with the estimation from a single fidelity, 
according to Eq.~(\ref{singlefid}). 

For example, the $\ket{\psi_i}$ may be a graph state 
basis, where the fidelities have been determined from 
the expectation 
values of the stabilizer operators \cite{kiesel}. In 
Section V we will discuss an example for four-qubit 
states.

\section{Examples}

In this section, we present several examples for the presented method 
and compare it with other estimation methods as well as with exact 
values of the entanglement measures. A Mathematica file with the used 
algorithms for the calculation of the Legendre transforms is available 
from the authors.

\subsection{Concurrence for isotropic states}
As a first example, let us consider isotropic states in a 
$N \times N$ system, defined by
\be
\vr(F)= \frac{1-F}{N^2-1}(\eins-\ketbra{\phi})+F\ketbra{\phi},
\label{isotropic}
\ee
which are a convex combination of a maximally entangled state 
$\ket{\phi}=\sum_i \ket{ii}/\sqrt{N}$ and the totally mixed state.
The parameter $F$ encodes the fidelity of $\ket{\phi},$ i.e., 
$F=\bra{\phi}\vr(F)\ket{\phi}.$ For these states, the concurrence
is known to be \cite{rungta}
\be
C(\vr)=\sqrt{\frac{2N}{N-1}}(F-\frac{1}{N}).
\ee
In order to test our methods, we consider the standard 
witness for states of the for states of the from in 
Eq.~(\ref{isotropic}), namely
\be
\WW = \frac{\eins}{N} - \ketbra{\phi},
\label{isotropicwitness}
\ee
and estimate from its expectation value the concurrence, 
using our algorithm. The results for the case $N=3$ are shown 
in Fig.~(2). It turns out that for this case, our lower bounds 
are sharp and reproduce the exact value of the concurrence.

\begin{figure}[t!!]
\includegraphics[width=0.99\columnwidth]{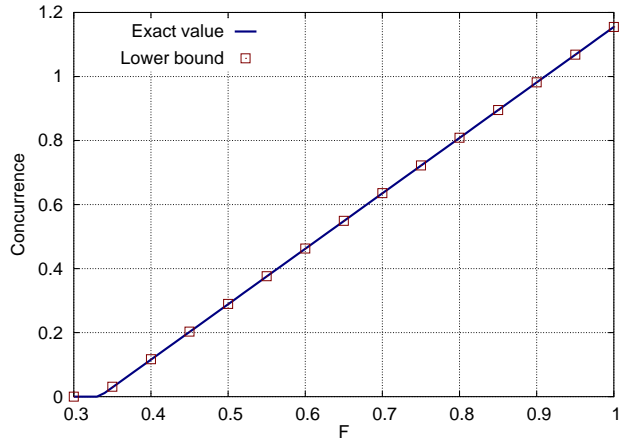}
\caption{Exact value of the concurrence and a lower bound 
from the witness in Eq.~(\ref{isotropicwitness}) and the 
algorithm in Sec. IV for isotropic states. See text for details.
}
\end{figure}

\subsection{Comparison with other estimation methods}

Let us compare the presented estimation method with other 
methods of estimating entanglement measures. For this aim, 
we consider two-qubit states of the form
\be
\varrho(p) = p \ketbra{\psi} + (1-p)\sigma,
\label{comparisonstate}
\ee
where $\ket{\psi}$ is a pure entangled state  and $\sigma$ is 
unknown and random separable noise. As entanglement measure
we consider the entanglement of formation. This is, for the 
case of two qubits, equivalent to the concurrence, since it is a 
monotonous function of it. We consider six different 
methods for estimating the entanglement of formation:

\begin{enumerate}

\item 
For the two-qubit case we can exactly calculate the 
entanglement of formation due to the formula of 
Wootters \cite{wootters1}. Clearly, since this requires 
complete knowledge of the density matrix, for an experimental 
implementation state tomography is needed.

\item In Refs.~\cite{cab1, cab2} a method to estimate the 
entanglement of formation or the concurrence from the separability 
criterion of the positivity of the partial transpose (PPT) 
or the computable cross norm or realignment criterion (CCNR) 
was presented. Experimentally, this approach requires 
again state tomography.

\item We can also take the witness which is proper for the 
states of the form $\vr(p) = p \ketbra{\psi}+ (1-p)\eins/4.$
This witness might not be the optimal one for the state under 
investigation, since the noise is not known and in general white. 
However, we can use the Legendre transform with the algorithm of 
Ref.~\cite{wir} to estimate the entanglement of formation from it.
Equivalently, we can use the algorithm of Section III to estimate
the concurrence.
Experimentally,  this method does not require state tomography, 
only three local measurements are needed for the measurement of 
the witness \cite{jmo}.

\item The witness in the third method is of the form
\be
\WW = \ketbra{\phi}^{T_B}.
\ee
Clearly, the mean value 
$\mean{\WW}=Tr(\vr_{\rm exp} \WW) = \bra{\phi}\vr_{\rm exp}^{T_B}\ket{\phi}$
can be used to derive a lower bound on the negativity of the 
partial transpose $\vr_{\rm exp}^{T_B}.$ Then, the second 
method \cite{cab1, cab2} may be used to estimate the 
entanglement of formation. This approach does not require 
state tomography.

\begin{figure}[t]
\includegraphics[width=0.99\columnwidth]{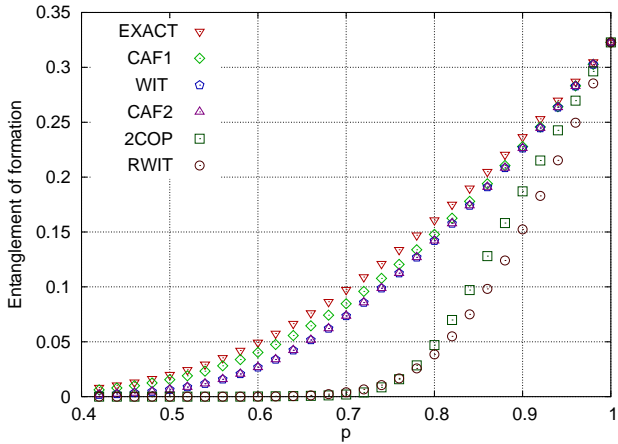}
\caption{Comparison of different methods to estimate the 
entanglement of formation for noisy two-qubit states. EXACT
denotes the exact calculation according to Ref.~\cite{wootters1}; 
CAF1 the second method, estimating the entanglement form the PPT criterion 
\cite{cab1,cab2}; WIT the estimation from the witness as presented in this 
paper and in Ref.~\cite{wir}; CAF2 the fourth method as a combination of the witness 
with CAF1; 2COP the fifth method, using measurements on 
two copies \cite{mintert1} and RWIT the sixth method \cite{mintert2}. 
See text for details.
}
\end{figure}

\item In Ref.~\cite{mintert1} lower bounds on the concurrence 
from measurements on two copies of the state $\vr$ have been 
derived.
For the case of two qubits we can use them to bound 
also the entanglement of formation. A measurement of a single
observable on two copies can always be expressed as a function 
of mean values of local observables on a single copy. In this 
case, however, for such an implementation effectively state 
tomography is needed.

\item In Ref.~\cite{mintert2} a method for estimating the 
concurrence from special entanglement witnesses has been 
derived. Namely, it was shown that
\be
E_C(\vr) \geq - \frac{1}{E_C(\ket{\phi})} \mean{\WW_\phi},
\label{redwit}
\ee
where $\WW_\phi= 2 [\eins_ A \otimes Tr_A(\ketbra{\phi}) - \ketbra{\phi}]$
is a witness belonging to the reduction criterion.
We use this method with the witness $W_\psi$ for the state 
$\ket{\psi}.$ With this choice, the method automatically 
reproduces the exact value for the case $p=1$ and it can be 
seen that Eq.~(\ref{redwit}) is nothing but the Legendre 
transform for a special choice of the slope $\lambda.$ Experimentally, 
this method would also require three local measurements \cite{mintert2,jmo}.

\end{enumerate}

As an example, we considered in Eq.~(\ref{comparisonstate})
the pure state $\ket{\psi}=(4\ket{00}+\ket{11})/\sqrt{17}$
and the state $\sigma$ was randomly (in the Hilbert-Schmidt 
measure) chosen from the set of separable states \cite{zs}. For a 
fixed $\sigma$ we calculated all the above mentioned values 
depending on the noise level $p$ and finally averaged over 
hundred realizations of $\sigma.$ The results are shown in 
Fig.~(3) and Table 1.
\begin{table}
\begin{tabular}{|l|c|c|c|c|c|}
\hline
Method &CAF1 & WIT & CAF2& 2COP & RWIT
\\
\hline
\# of measurements & 9 &  3 & 3 & 9 & 3
\\
\hline
$\eta$ for $p\in [0.8; 1.0]$ & 96.0 \% & 94.8 \% & 94.8 \% &70.1 \% & 63.0 \%
\\
\hline
$\eta$ for $p\in [0.6; 0.8]$ & 86.0 \% & 72.0 \% &  72.0 \% & 5.9 \%& 7.0 \%
\\
\hline
\end{tabular}
\caption{Comparison of the different estimation methods. 
For each method, the required number of local measurement 
settings on a single copy and the efficiency $\eta$ is 
given. Here, the 
efficiency is defined as the  ratio between the estimated
value of the entanglement of formation and the actual one, 
for two different regions of the noise parameter $p.$
} 
\end{table}

One can clearly see that the methods two, three and four result 
in bounds on the entanglement of formation, which are very close 
to the exact result. The second method is the best bound, however,
it requires complete knowledge of the state. The third method is 
by construction better than the fourth method (as the Legendre 
transform delivers by construction the best possible bounds from 
a given witness). In this example, however, they are practically 
equivalent. The fifth and sixth method deliver good results if 
$\vr$ is close to a pure state.

\subsection{Geometric measure for four-qubit states}

As a third example we discuss the geometric measure 
of entanglement. In order to demonstrate the method
in Section IV, we consider an experimental situation 
similar to the one in Ref.~\cite{kiesel}. In this 
experiment, a four-photon cluster state 
\be
\ket{CL}=\frac{1}{2}(\ket{0000}+\ket{0011}+\ket{1100}-\ket{1111})
\ee
has been prepared using parametric down conversion and a 
controlled phase gate. The fidelity of the target state
was then determined by the measurement of all stabilizer 
operators. These operators are given by the local observables 
$S_1=\sigma_z \sigma_z \eins \eins$,
$S_2=\sigma_x \sigma_x \sigma_z \eins$,
$S_3=\eins \sigma_z \sigma_x \sigma_x$,
and
$S_4= \eins \eins \sigma_z \sigma_z$,
and products of these observables. The cluster state is 
an eigenstate (with eigenvalue $+1$) of all these $2^4=16$ 
observables, and from their expectation values the 
fidelity can be determined \cite{kiesel, tothguehne}.

Using the fact that the maximal overlap of the cluster
state with fully separable states equals $1/4$ (i.e., 
the geometric measure is $3/4$ \cite{markham1}) 
Eq.~(\ref{singlefid})
can be used to bound $E_G$ from this fidelity. There are, 
however, also other common eigenstates of the $S_i$ with
different eigenvalues. These states are orthogonal to the 
cluster state and form the so-called cluster state basis.
All states in this basis share the same entanglement 
properties and their 
fidelities can also be determined from the mean values 
of the $S_i.$ In fact, the knowledge of all $\mean{S_i}$
is equivalent to the knowledge of all fidelities.

In order to investigate how the information about the
fidelities of all states in the cluster state basis can 
be used for the estimation of entanglement, we consider the 
simple case that only three fidelities are larger 
than zero, $F_1$, $F_2$ and $F_3 = 1-F_1-F_2.$ 
For a given triple of fidelities we may first consider 
the maximal fidelity and then use Eq.~(\ref{singlefid})
to obtain a lower bound on $E_G.$ Alternatively, we can 
use the methods of Section IV and consider all three 
fidelities at the same time. In practice, this gives a 
lower bound on $E_G$ by the optimization problem
\be
E_G \geq \sup_{\lambda_1,..., \lambda_4}
\{
\sum_{k=1}^3 \lambda_k F_k - 
\big\Vert
\left[
X
\right]
\big\Vert
+1
\},
\ee
where $X$ is defined as in Eq.~(\ref{anageo}) for $k=4.$
Any set of $\lambda_i$ delivers already a valid lower bound, 
and the optimum  over all $\lambda_i$ is easily found.

\begin{figure}[t]
\includegraphics[width=0.99\columnwidth]{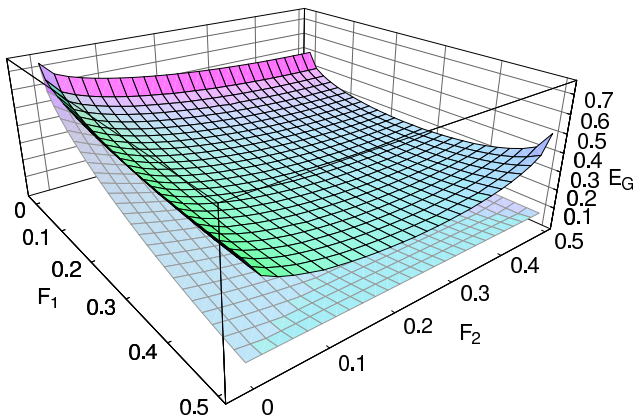}
\caption{Comparison between two estimation methods for 
the geometric measure of entanglement. The lower curve
is the analytical bound from the maximal fidelity according
to Ref.~\cite{wir} or Eq.~(\ref{singlefid}), the upper 
curve is the analytical bound from Section IV, taking all fidelities 
into account. See text for details. 
}
\end{figure}

The results are plotted in Fig.~(4). One can clearly see, that
taking all fidelities into account, improves the lower bounds
significantly.

\section{Conclusion}
In conclusion, we have investigated how entanglement measures can 
be estimated from incomplete experimental data. We have shown that
the method of Legendre transforms can successfully be applied to 
the concurrence and extensions thereof. Furthermore, we have 
presented an analytical way to estimate the geometric measure
if the fidelities of certain basis states are known. Extending 
the presented methods to other entanglement measures is an 
interesting task  for further study. 

We thank J. Eisert, N. Kiesel, A. Osterloh, J. Siewert and  
W. Wieczorek for discussions. This work has been supported 
by the FWF, the DFG, and the EU (OLAQUI, PROSECCO, QUPRODIS, 
QICS, SCALA).


\begin{thebibliography}{45}
\bibitem{remark1} Here, we also use the conditions
that $E$ is continuous, and that the state space is compact. 
These conditions may be relaxed by requiring only that
the sublevel sets $s_t=\{\vr | E(\vr) \leq t\}$ are compact.
See Ref.~\cite{reimpell} for a more detailed discussion.
\end{thebibliography}
\end{document}